\documentclass[preprint2,longabstract]{aastex}

\input{psfig}

\def\farcs{\hbox{$.\!\!^{\prime\prime}$}}

\def\kms      {\ifmmode{\rm km\,s}^{-1} \else km\,s$^{-1}$\fi}
\def\mujybm{\ifmmode{\rm \mu Jy}\,{\rm beam}^{-1}\else${\rm \mu}$Jy\,beam$^{-1}$\fi}
\def\ltsim{\ifmmode\stackrel{<}{_{\sim}}\else$\stackrel{<}{_{\sim}}$\fi}
\def\gtsim{\ifmmode\stackrel{>}{_{\sim}}\else$\stackrel{>}{_{\sim}}$\fi}

\shorttitle{Prompt radio emission from supernova 2004dj}
\shortauthors{Beswick et al.}

\begin{document}

\title{Monitoring of the prompt radio emission from the unusual supernova 2004dj in
NGC\,2403}

\author{R. J. Beswick$^1$, T. W. B. Muxlow$^1$, M. K. Argo$^1$,
A. Pedlar$^1$, J. M. Marcaide$^2$ and~K.~A.~Wills$^3$}
\affil{1)  Jodrell~Bank Observatory, The University of Manchester,
Macclesfield, Cheshire, SK11~9DL, UK (rbeswick@jb.man.ac.uk; twbm@jb.man.ac.uk; mkargo@jb.man.ac.uk; ap@jb.man.ac.uk.\\2) Departamento de Astronom\'{\i}a,
Universitat de Val{\`e}ncia, 46100 Burjassot, Spain (J.M.Marcaide@uv.es)\\3) Department of
Physics \& Astronomy, University of Sheffield, Sheffield, S3 7RH, UK. (K.Wills@sheffield.ac.uk)}

\begin{abstract}

Supernova 2004dj in the nearby spiral galaxy NGC\,2403 was detected optically in July 
2004.  Peaking at a magnitude of 11.2, this is the brightest supernova
detected for several years.  Here we present Multi-Element Radio Linked Interferometer Network 
(MERLIN) observations of this source, made over a four month period, which give a 
position of R.A. = 07$^{\rm h}$37$^{\rm m}$17.044$^{\rm s}$, Dec =
+65$^{\rm o}$35\arcmin57\farcs84 (J2000.0).  We also present a
well-sampled 5\,GHz light curve covering the period from 5 August to 2 December 2004.
With the exception of the unusual and very close SN 1987A, these
observations represent the first detailed radio light curve for the
prompt emission from a Type II-P supernova. 

\end{abstract}


%--------------------------------------------------------------------------------------

\section{Introduction}

We are currently using MERLIN\footnote{MERLIN is a national
facility operated by the University of Manchester on behalf of PPARC
in the UK.} (Multi-Element Radio Linked Interferometer Network) and NRAO's VLA\footnote{The National Radio Astronomy Observatory is a facility of the National Science Foundation operated under cooperative agreement by Associated Universities, Inc} (Very Large Array) to monitor the radio emission associated with a sample of
nearby starburst galaxies \cite{argo04a} with the objective of determining their radio
supernova (RSn) rates.  As the active star-forming regions of these galaxies contain 
large numbers of massive stars, it is likely that most of the RSn will be associated 
with core-collapse supernovae (Type Ib/c and Type II).  In addition to regular monitoring 
of the galaxies in our sample, we plan to use MERLIN to monitor the radio flux-density of 
other nearby supernovae events where possible (e.g. Beswick et al. 2004a,b)
\nocite{beswick04a,beswick04b}, particularly when 
the VLA is in compact configurations.

The detection of SN 2004dj, visually on 31 July 2004 \cite{nakano04} and
in the radio using the VLA at 8.4 GHz on 2 August \cite{stockdale04},
provided an ideal opportunity to study a nearby supernova and its
evolution at radio wavelengths.  The high declination of NGC\,2403, the
host galaxy (+65\degr), makes it an ideal source for high resolution
MERLIN observations.

Peaking at magnitude 11.2, SN\,2004dj was the brightest optical supernova
for several years.  Shortly after the initial detection it was reported
that the spectrum of SN\,2004dj showed features typical of Type II-P
supernovae \cite{patat04}. Type II-P supernovae are believed to originate
via core-collapse in hydrogen rich, massive stars.  The main
characteristic of Type II-P supernovae is that their optical luminosity,
unlike other Type II supernovae, does not decay rapidly after peaking but
remains approximately constant for two to three months before decaying
exponentially.  The physics of supernovae are discussed in detail in
Woosley \& Weaver (1986)\nocite{woosley86} and references therein.

Despite the fact that Type II-P supernovae are relatively common
optically, few have been detected in the radio.  As far as we are aware,
only two Type II-P supernovae have been detected at radio wavelengths: SN
1987A \citep[][and~refs.~therein]{ball95,kirk94,turtle87} and SN 1999em
\cite{pooley02}.  A further possibility is SN 1923A, although this is
arguably a supernova remnant rather than a radio supernova \cite{eck98}.
The only detailed radio study of a Type II-P RSn is SN 1987A (see section
\ref{87A}), whereas the radio observations of SN 1999em are rather limited
with relatively few detections between day 34 and 70, despite extensive
monitoring between 8.4\,GHz and 1.4\,GHz.  Consequently no well-sampled
radio light curve of SN 1999em was established.

SN 2004dj occurred in NGC\,2403, a nearby (3.2\,Mpc, Karachentsev et al.
2004\nocite{karachentsev04}) spiral galaxy located within the M81 group.  
As this is approximately half the distance of SN 1999em, this offered an
excellent opportunity to establish a Type II-P radio light curve, so using
a sub-set of the MERLIN array, we began monitoring this source frequently
at 5GHz on August 5th and continued through to December 2nd 2004.

%%%%%%%%%%%%%%%%%%%%%%%%%%%%%%%%%%%%%%%%%%%%

\section{Observations}

Monitoring observations at 4994\,MHz with a bandwidth of 14.75\,MHz were
made with a sub-set of the MERLIN array \cite{thomasson86} between 5th
August and 6th October 2004.  These employed a single 218\,km baseline
between the Cambridge and Defford antennas. Primary flux calibration was
performed using the strong source 3C84 assuming a flux density of
14.53\,Jy, which was derived via a comparison with 3C286. The nearby point
source 0733+646 was used for primary phase calibration and for secondary
flux calibration assuming a flux density of 0.3566\,Jy.  Two additional
secondary phase calibration sources (0713+669 and 0752+639) were also
observed in order to establish the precision to which point source
positions could be derived from single baseline data.  Tests using all
three phase calibration sources have shown that we are able to derive the
position of SN 2004dj to better than 50 mas.

\begin{figure*}
%\psfig{figure=f1.eps,width=8cm,angle=0}
% single column (small) -- remove *s
\psfig{figure=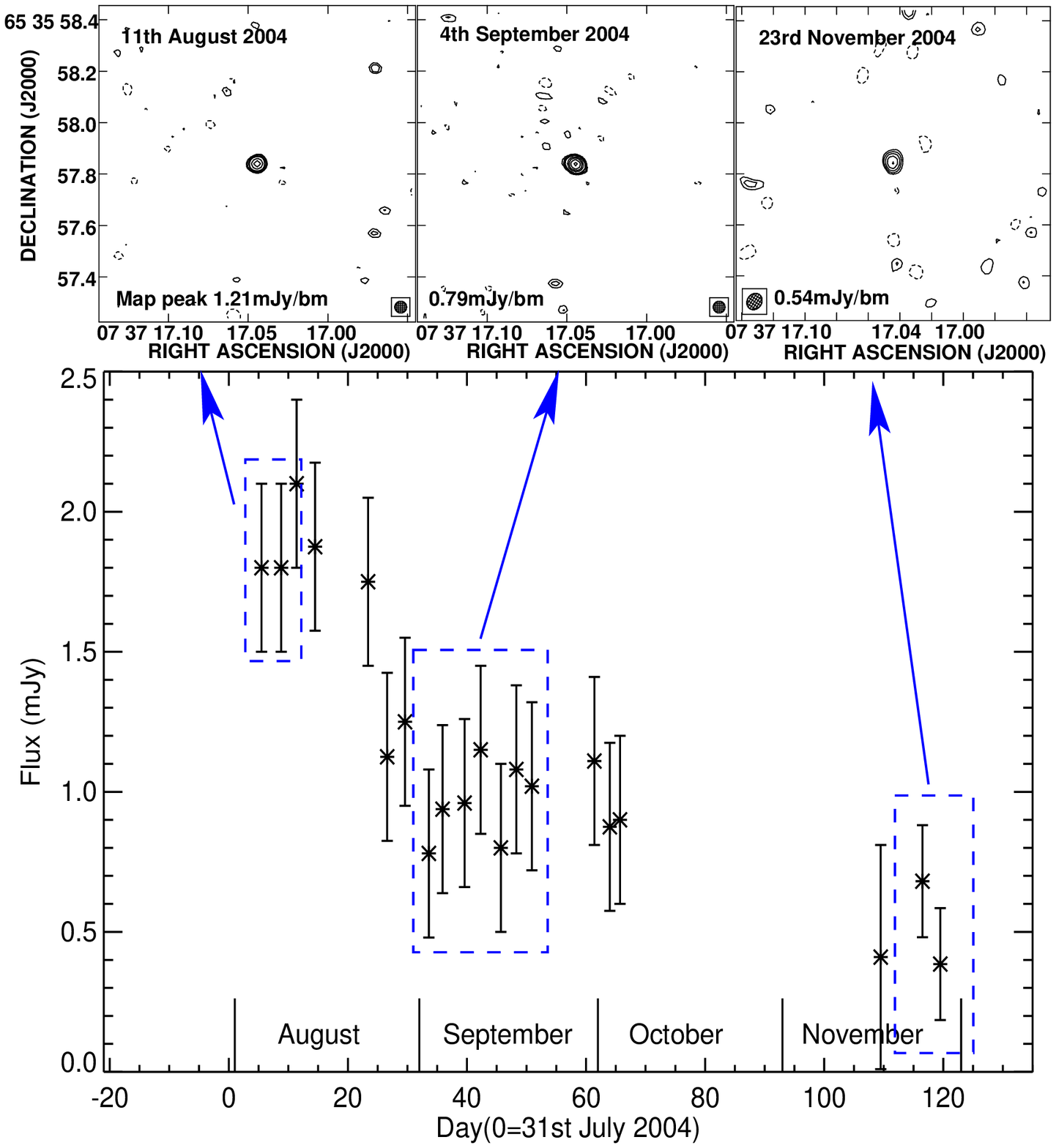,width=16cm,angle=0}
% Double column (BIG) -- with *s
\caption{\label{rcurve}Radio light curve of SN 2004dj from MERLIN
data. Day 0 of the radio light curve is 31st July 2004. The three
upper panels show contoured images of SN 2004dj at different
stages of its evolution. Observations up until October were performed
using a single 217\,km baseline, following this the entire MERLIN array
was used. Each image is contoured with multiples of
$\sqrt2$ times 165, 63 and 132\,$\mu$Jy\,beam$^{-1}$, from left to right
respectively. The convolved beam size in the three images is
45$\times$45\,mas, 45$\times$45\,mas 58$\times$57\,mas respectively. The peak flux of the source at each epoch is shown in
the bottom left-hand corner of the images. The dates shown on each of
the three images represent the start date for each of the imaged epochs.}
\end{figure*}

The single-baseline `maps', shown in Fig.\,\ref{rcurve}, illustrate that
SN\,2004dj is detected at greater than 3$\sigma$ in each of our 2.5 day
integrations and that we are able to measure its position from such
images. It should be noted that, although these single-baseline `maps' can
only provide limited structural information regarding the source, its
position and flux density can be obtained. The light curve from August to
early October (Fig.\,\ref{rcurve}) is derived from 2.5 day vector averaged
points after rotating the data from the nominal pointing position to the
measured position for SN 2004dj. Additional points on the light curve
(November onward) are derived from observations made using the full MERLIN
array and were calibrated and imaged using standard methods.

%%%%%%%%%%%%%%%%%%%%%%%%%%%%%%%%%%%%%%%%%%%%

\section{Results \& Discussion}

\subsection{Position and progenitor}

From these MERLIN observations, the position of SN 2004dj is found to be
R.A. = 07$^{\rm h}$37$^{\rm m}$17$^{\rm s}\!\!$.044, Dec. =
+65\degr35\arcmin57\farcs84 (J2000.0), with an absolute error of
$<$50\,mas. No proper motion is detected between any of our epochs. This
position is within 0.4 arcsec of the optical position reported by Nakano
et al. (2004)\nocite{nakano04} and agrees with the $Chandra$ X-ray
position reported by Pooley \& Lewin (2004)\nocite{pooley04}.  However,
the VLA D-configuration position reported by Stockdale et al. (2004)  
(R.A. = 7$^{\rm h}$37$^{\rm m}$16$^{\rm s}\!\!$.916, Dec =
$+$65\degr35\arcmin56\farcs97) is $\sim$1.2\,arcsec away from the MERLIN,
{\it Chandra} and optical positions. The error in this radio position is
almost certainly the result of confusion from the extended radio emission
from NGC\,2403 \cite[][ S.\,D.\,Van~Dyk, private
communication]{filippenko04_8391}.
 
This position measurement puts the supernova near to n2403-2866 (R.A. = $07^{\rm
h}37^{\rm m}16^{\rm s}$.93, Dec = $+65^{\rm o}$35' 57\farcs7; Larsen \& Richter
1999\nocite{larsen99}) a cluster also known as Sandage 96, and the
luminous blue variable (LBV) candidate 7-3909 (R.A. = 07$^{\rm h}$37$^{\rm m}$16$^{\rm s}$.93, Dec = 
+65$^{\rm o}$35\arcmin57\farcs6), another suggested progenitor \cite{weis04}.  Optical
observations suggest that the object is extended and therefore
more likely to be a cluster than a single LBV \cite{apellaniz04}.

\subsection{Radio light curve}

Using a single MERLIN baseline, the 5\,GHz flux of SN 2004dj has been monitored on a
regular basis from the 5th August through to the 6th October 2004.
Additional flux measurements using the full MERLIN array were made on several
occasions in late-November and early-December.

These radio observations, initiated just five days after the initial optical
detection by K. Itagaki \cite{nakano04}, represent the best sampled early radio light curve of a
Type II-P supernova made to date, with the obvious exception of SN 1987A. At each epoch
SN 2004dj was detected and imaged.  The radio light curve for SN 2004dj, derived from the 
MERLIN observations, is shown in Fig.\,\ref{rcurve}.

On 5th August the prompt radio emission from SN 2004dj was detected with a
flux density of 1.8\,mJy at 5\,GHz \cite{argo04b}.  Following this initial
detection, the flux of SN 2004dj was observed to reduce during the second
half of August before reaching an approximately constant flux value of
$\sim$0.9\,mJy. The flux of SN 2004dj showed
little significant variation from 0.9$\pm$0.4\,mJy over the
following month before decaying slowly from this radio plateau level.
By December (day $\sim$125) the flux had decreased to the point where the supernova
was undetectable above the noise ($\sigma_{rms} \sim
0.1$\,mJy\,bm$^{-1}$, December map, not presented here).  It is not known when the 6\,cm
peak occurred, but it is possible that the earliest of our observations coincided with, 
or closely followed, the peak radio flux at 5\,GHz; placing limits on both the timing and 
flux of the peak of the prompt radio emission.

If it is assumed that our initial observations, in early August, do
coincide with the 5\,GHz peak in the prompt radio emission from SN 2004dj, 
this implies a peak flux of S$_{\rm
5\,GHz}=1.9\pm0.1$\,mJy\,bm$^{-1}$, equivalent to  L$_{\rm
6cm\,peak}\approx2.45\times10^{18}$\,W\,Hz$^{-1}$. At this peak
radio luminosity, SN 2004dj is, along with the Type II-P supernova 1999em,
one of the least luminous radio supernova detected (with the
exception of the unusual LMC supernova 1987A, which was about
two orders of magnitude less luminous). 

Based upon radio observations of several Type II supernova, Weiler et
al.\,(2002) derive an empirical relationship between peak luminosity
and time delay between explosion and its peak such that,

\begin{eqnarray}
 {\rm L}_{\rm 6cm\,peak} &\!\!\ \simeq &\!\!\! 5.5^{+8.7}\!\!\!\!\!\!\!\!\!_{-3.4}\times10^{16} \nonumber\\
	  &\!\!\! \times &\!\!\! (t_{\rm 6cm\,peak} - t_0)^{1.4\pm0.2}\,{\rm W}\,{\rm Hz}^{-1}
\end{eqnarray}

where $t_{\rm 6cm\,peak} - t_0$ is the number of days between the
supernova detonation and the peak of the prompt radio emission at
6\,cm. Assuming the peak luminosity of SN 2004dj at 6\,cm is
$2.45\times10^{18}(\pm 10\%)\,{\rm W}\,{\rm Hz}^{-1} $ this implies $t_{\rm
6cm\,peak} - t_{0} = 15^{+42}\!\!\!\!\!\!\!\!_{-10}$ days, placing
the detonation date between July 11th and the first detection of
SN\,2004dj, July 31st. This is broadly consistent with spectroscopic observations of
SN\,2004dj, obtained on August 3rd \cite{patat04}, which report SN 2004dj to
have a spectrum  similar to that of SN 1999em about three weeks after
explosion (i.e. around July 13th). Supernova 2004dj is a Type II-P
supernova and only the second of this Type for which a well sampled
radio light curve exists. Thus, although these observations appear to be
approximately consistent with equation\,1, more observations of the prompt radio emission
from Type II-P supernovae are required to make conclusions regarding the
whole of this class of supernovae, if their radio emission is to be
used as a reliable luminosity distance indicator.

\subsection{\label{87A}Comparison with SN 1987A and SN 1999em}

The only other Type II-P supernova to have been extensively monitored
and imaged at radio wavelengths in its early evolution is the
supernova 1987A \cite[][and
refs. therein]{ball95,kirk94,turtle87}. By virtue of the proximity of SN 1987A, situated 
in the LMC at a distance of
$\sim$50\,kpc, the prompt radio emission from this supernova was
detected a couple of days after its optical detection and has been
monitored at various frequencies ever since \cite{turtle87}.  

The peak luminosity of the prompt radio emission from SN 1987A at GHz frequencies was
4$\times10^{16}$\,W\,Hz$^{-1}$ \cite{turtle87}, approximately two
orders of magnitude lower than the peak luminosity at 5\,GHz recorded for
SN\,2004dj ($\gtsim2.6\times10^{18}$\,W\,Hz$^{-1}$). However, the 5\,GHz peak luminosity
of SN 2004dj is comparable to that of Type II-P supernova 1999em which was estimated to have a L$_{\rm
6\,cm\,peak}\sim2.2\times10^{18}$\,W\,Hz$^{-1}$ \cite{pooley02}.
The X-ray luminosity of SN 2004dj is approximately three times greater than that of SN 
1999em \cite{pooley04}.  

Our estimate of  $t_{\rm 6cm\,peak} - t_{0} =
15^{+42}\!\!\!\!\!\!\!\!_{-10}$\,days, and the spectroscopically based
estimates of $\sim$30\,days of Patat et al. (2004), are comparable to
the estimates for SN 1999em ($\sim$34\,days) \cite{pooley02}. In both of
these cases, the time delay before the peak of the supernova's GHz
radio emission is an order of magnitude longer than that observed for
the Type II-P supernova 1987A. 

Along with the detection of prompt radio emission from supernovae 
1987A, 1999em and 2004dj, all three of these sources have also been
detected at X-ray wavelengths \cite{masai87,dotani87,rodin87,pooley02,pooley04}.
This is often true for core collapse 
supernovae (see figure\,1 of Immler \& Lewin 2003) and is unsurprising since both 
types of emission arise from the shocked circumstellar medium. Consequently, it
appears that an X-ray detection of a Type II-P supernova may be
an indicator that it can also be detected at radio wavelengths, or vice
versa. However, to date very few supernovae of this type have been
detected in either of these wavebands.

Interestingly, in the only other case where a Type II-P supernova has
been extensively monitored at radio wavelengths, SN 1987A,  the radio
flux was found to evolve and brighten after $\sim$1400 days
\cite[][and refs. therein]{manchester02}, signifying the birth of a radio
remnant. In fact, since this
re-detection of SN 1987A, its radio emission has
continued to increase at an approximately monotonic rate and
now far exceeds the radio brightness of its initial prompt radio
emission phase. This increase in the observed radio luminosity of SN 1987A 
occured as the expanding shock front from the supernova encountered a large increase in 
the density of the circumstellar medium.  Coincidently, this second phase of emission 
was also observable at X-ray wavelengths.  Uniquely, SN 2004dj's proximity may provide a 
second opportunity to study the transition of a radio supernovae into a
supernova remnant, although this will be dependent upon the density of the surrounding 
circumstellar medium, and thus the nature of the progenitor. Additionally if, in a few 
years time, SN 2004dj does
significantly brighten, it may be detectable and resolvable using high sensitivity
VLBI observations. As such it is essential that continued radio
monitoring observations of this source are made.

\acknowledgments

We would like to thank all the MERLIN staff for their assistance in making these
observations possible. MKA acknowledges support from a PPARC
studentship. RJB would like to acknowledge financial support by the
European Commission's I3 Programme ``RADIONET'' under contract No.505818.

\clearpage

\end{document}